\documentclass[showpacs,preprintnumbers,amsmath,amssymb,nofootinbib]{revtex4}
\usepackage{fontenc}
\usepackage{graphicx}

\usepackage{amsmath}
\begin{document}
\title{Greybody factors for topological massless black holes}
\author{Pablo Gonz\'{a}lez}
\email{pablo.gonzalezm@mail.udp.cl} \affiliation{Instituto de F\'{\i}sica,
Pontificia Universidad Cat\'olica de Valpara\'{\i}so, Casilla
4950, Valpara\'{\i}so, Chile} \affiliation{Universidad Diego Portales, Casilla 298-V, Santiago, Chile.}
\author{Joel Saavedra}
\email{joel.saavedra@ucv.cl} \affiliation{Instituto de
F\'{\i}sica, Pontificia Universidad Cat\'olica de Valpara\'{\i}so,
Casilla 4950, Valpara\'{\i}so, Chile.}
\author{Cuauhtemoc Campuzano}
\email{ccvargas@uv.mx} \affiliation{Departamento de F\'\i sica,
Facultad de F\'\i sica e Inteligencia Artificial, Universidad
Veracruzana, 91000, Xalapa Veracruz, M\'exico}
\author{Efra\'\i n Rojas}
\email{efrojas@uv.mx} \affiliation{Departamento de F\'\i sica,
Facultad de F\'\i sica e Inteligencia Artificial, Universidad
Veracruzana, 91000, Xalapa Veracruz, M\'exico}
\date{\today}
\begin{abstract}
We study the greybody factors, the reflection and transmission coefficients for a non-minimally coupled massive scalar field in a $d$-dimensional topological  massless black hole background in the zero-frequency limit. We show that there is a range of modes contributing to the absorption cross section, contrary to the current results where the mode with lowest angular momentum contributes alone to the absorption cross section.
\end{abstract}
\maketitle
\section{Introduction}

The Hawking radiation is an important
quantum effect in black hole physics but, this one is enigmatic because at quantum
level black holes are not 'black' completely since these emit radiation
with a temperature given by $h/8 \pi k_{B}GM$ \cite{Hawking:1974sw},
contrary to the classical context where it is believed that anything
can escape from them.
The originated thermal radiation at the black hole event horizon is emitted
into surrounding spacetime  with the consequence that the semiclassical approach
for a black hole exhibit that it slowly loose its mass and eventually
evaporates. At the event horizon the Hawking
radiation is in fact blackbody radiation. However, this radiation still
has to traverse a non-trivial curved spacetime geometry before it
reaches after all to an observer where it is detected. The surrounding spacetime thus works as a potential barrier for the
radiation giving a deviation from the blackbody radiation spectrum,
as it is detected by an asymptotic observer \cite{Harmark:2007jy,Maldacena:1996ix}.

The greybody factors are the probabilities for outgoing waves in the
$\omega$-mode to reach infinity.
the horizon which filter the initially blackbody spectrum emanating from the horizon
\cite{Maldacena:1996ix}.
If one integrates the greybody factors
over all spectra, the total black hole emission rate is obtained. Moreover,
if they are constant the black hole emission spectrum would be exactly that
of a blackbody radiation. This is the non-triviality of the greybody factor
which leads to deviations of blackbody emissions and the consequent
greybody radiation \cite{Harmark:2007jy}. Essentially, such radiation
possesses a thermal character and inevitably black holes slowly
evaporate. Furthermore, in the understanding of the event horizon for a
black hole, the Hawking radiation plays an important role providing
clues about the quantum structure of GR. In order to study the Hawking
radiation we need to allow quanta to fall into the hole. The absorption
cross section for low energy particles in $3+1$ dimensions consider a
particle to be a massless minimally coupled scalar field.
The cross section equals the
area of the black hole \cite{staro,gibbons75,page,unruh}. Additionally, it was shown in \cite{Das:1996we} that for all spherically
symmetric black holes the low energy cross section for massless
minimally coupled scalar fields is always the area of the horizon.
Nevertheless, contrary to the existing results, we show in this work that there is
a range of modes which contribute to the absorption cross section in the zero-frequency limit.

A well known fact to be mentioned that in four dimensions the Einstein tensor is the only symmetric and
conserved tensor depending on the metric and its derivatives, which
is linear in the second derivatives of the metric. The invariant
action that gives rise to these fields equations is the Einstein-Hilbert action
with cosmological constant $\lambda$. Alike, in higher dimensions
the Lanczos-Lovelock (LL) action \cite{LL} which is non-linear in
the Riemann tensor, gives rise also to second-order field equations. Hence, for describing higher dimensional black
holes this type of action results useful. An exhaustive analysis for
asymptotically local AdS black holes geometries for nontrivial topologies of the
tranverse section was performed in \cite{Aros:2000ij}.

The purpose of this work is to compute the greybody factors, the reflection
and transmission coefficients for
topological massless black holes with nontrivial topology of the
transverse section asymptotically AdS in $d$ dimensions.
The paper is organized as follows. In Sec. II we provide
mathematical preliminaries and we describe the background spacetime
that we will use along the work. In Sec. III, we study the scalar
perturbation of $d$-dimensional topological massless black holes. We
find the associated greybody factors and the reflection and transmission
coefficients in Sec. IV. Also, in this section we evaluate numerically our results,
specialized to the $4$-dimensional case. We finish with some comments and
discuss the relevance of our results in Sec. V.
\section{Gravity in higher dimensions}

The Lanczos-Lovelock (LL) action is the outstanding extension of
general relativity in $d$-dimensional space-times that leads to
second order field equations for the metric \cite{LL}. It is given
by
\begin{equation}
\label{dtopologicalaction}\
S_{\mbox{\tiny LL}}[g_{\mu \nu}]
=
\kappa \int \sum_{p=0}^{k}c_{p}^{k}L^{p}~,
\end{equation}%
where $
L^{p}=\epsilon_{\alpha_{1}...\alpha_{d}}R^{\alpha_{1}\alpha_{2}}...R^{\alpha_{2p-1}
\alpha_{2p}}e^{\alpha_{2p+1}}...e^{\alpha_{d}}$, and $e^{\alpha}$ and $R^{\alpha\beta}$
stand for the vielbein and the curvature two-form ($\alpha, \beta = 0,1,\ldots,d-1$),
and $c_{p}^{k}=\frac{l^{2(p-k)}}{d-2p}(^{k}_{p})$ for $p\leq k$
and it vanishes for $p>k$, with $1\leq k\leq [\frac{d-1}{2}]$ ($[x]$
denotes integer part of $x$).
The constants $\kappa$ and $l$, are related to the gravitational constant $G_{k}$
and the cosmological constant $\Lambda$ through
\begin{eqnarray}\label{definitionk}\
\kappa&=&\frac{1}{2(d-2)!\Omega _{d-2}G_{k}}~,
\\
\label{lambda}\
\Lambda&=&-\frac{(d-1)(d-2)}{2l^2}~,
\end{eqnarray}%
where  $\Omega _{d-2}$ corresponds to the volume of the $(d-2)$-dimensional sphere.

The static black hole-like geometries possessing topologically non-trivial AdS
asymptotic behaviors admitting a unique global 
vacuum were found in
\cite{Aros:2000ij}.
These theories and their corresponding solutions were classified
by the integer $k$ which corresponds to the highest power of
curvature into the LL Lagrangian. Such solutions describe a
non-trivial $(d-2)$-dimensional transverse spatial section, $\sum_{\gamma}$.
These surfaces are labelled by the constant $\gamma=+1, -1, 0$,
depending on the curvature of the transverse section associated to
a spherical, hyperbolic or plane section, respectively. The set of
solutions describing a black hole in a free torsion theory, given
by \cite{Aros:2000ij}
\begin{equation}
\label{dtopologicalmetric}\
ds^{2}=-\left[ \gamma +\frac{r^{2}}{l^{2}}-\alpha \left( \frac{2 \mu G_{k} }{%
r^{d-2k-1}}\right) ^{\frac{1}{k}}\right] dt^{2} +
\frac{dr^{2}}{\left[ \gamma +%
\frac{r^{2}}{l^{2}}-\alpha \left( \frac{2 \mu G_{k}}{r^{d-2k-1}}\right) ^{%
\frac{1}{k}}\right] }+r^{2}d\sigma _{\gamma }^{2}~,
\end{equation}%
where $\alpha =(\pm 1)^{k+1}$ and the constant $\mu$ is related to the
black hole horizon $r_{+}$ through
\begin{equation}\label{relationr}\
\mu =\frac{r_{+}^{d-2k-1}}{2G_{k}}\left(\gamma
+\frac{r_{+}^{2}}{l^{2}}\right)^{k}~,
\end{equation}%
possesses an asymptotic behavior which is locally AdS for any
topology of $\Sigma_\gamma$. On the other hand, $\mu$ is also
related to the black hole mass $M$ by $
\mu =\frac{\Omega _{d-2}}{\Sigma _{d-2}}M+\frac{1}{2G_{k}}\delta
_{d-2k,\gamma }
$. Here, $\Sigma _{d-2}$ denotes the volume of the transverse
space. The conditions that the metric (\ref{dtopologicalmetric})
must fulfill in order to have an appropriate black hole solution
have been extensively discussed in \cite{Aros:2000ij,Aros:2002te}.

\section{Scalar perturbation for a $d$-dimensional topological massless black hole}

\begin{figure}[th]
\includegraphics[width=4.0in,angle=0,clip=true]{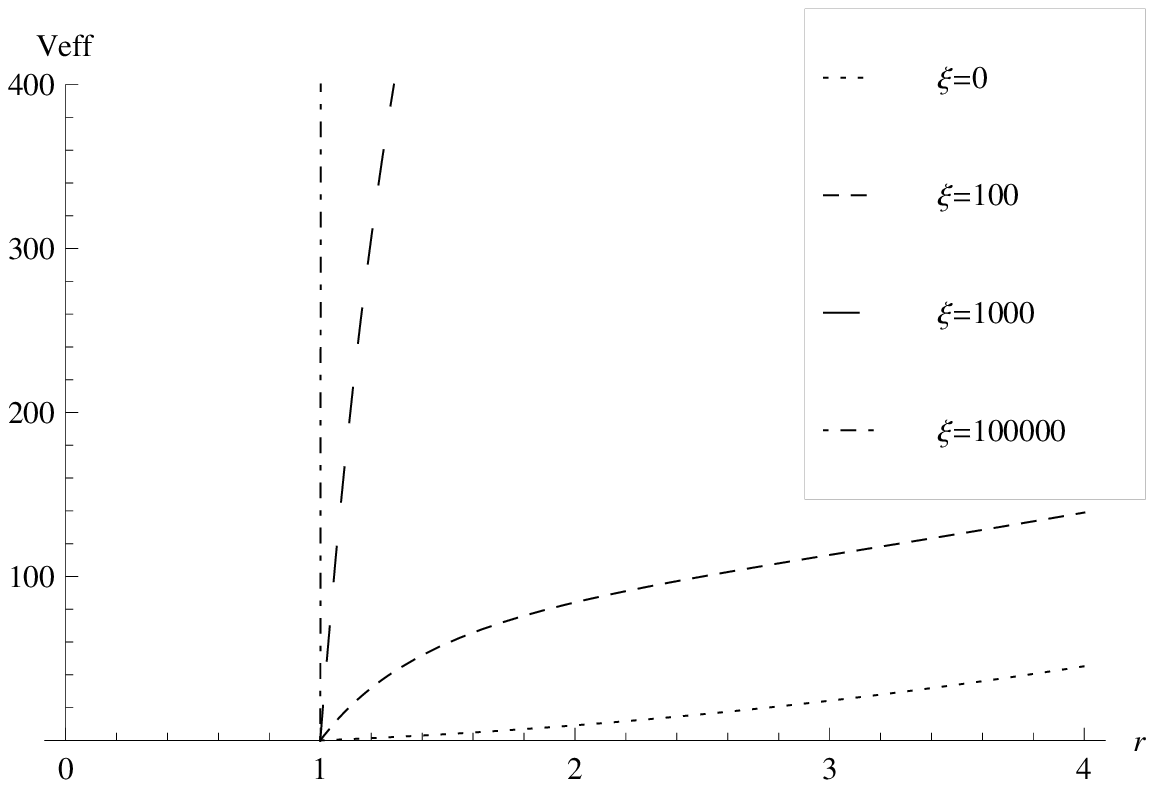}
\caption{$V_{\mbox{\tiny eff}}$ v/s r; $d=4$, $l=1$ and $m_{\mbox{\tiny eff}}=0$.}
\label{pot}
\end{figure}

If we specialize the solution (4) for $\mu=0$, the horizon geometry is described by a
negative constant curvature with $\gamma=-1$. In consequence the metric (4) reads
\begin{equation}\label{metric2}\
ds^{2}=-f(r)dt^{2}+\frac{1}{f(r)}dr^{2}+r^{2}d\sigma^2~,
\end{equation}
where $f(r)=-1 + \frac{r^{2}}{l^{2}}$ and $d\sigma^2$ is the line element of
a $(d-2)-$dimensional surface, $\Sigma_{d - 2}$.
Clearly, this metric has a horizon at $r_+ = l$.  It was shown in \cite{Aros:2000ij} that this solution not necessarily
describe a black hole. In fact, if the transverse section $\Sigma$ has the
topology $R^{d-2}$, the metric (\ref{metric2}) does not represent a black hole. It could be the case provided suitable identifications are performed on $\Sigma_{-1}$
\cite{Vanzo:1997gw,Brill:1997mf}. As mentioned in the Introduction, to gain insight into
the quantum nature of black holes the kinematical properties provide relevant
clues about their semiclassical aspects. In this spirit, the scalar perturbations
on a massless black hole are dictated by a massive non-minimally coupled scalar field,
$\phi$, propagating in the vicinity of the massless black hole. The action governing
the dynamics of the fields is
\begin{equation}
S[g_{\mu \nu},\phi]= S_{\mbox{\tiny LL}} + \int d^dx \sqrt{-g}
\left( \frac{1}{2} \partial^\mu \phi \partial_\mu \phi +
\frac{1}{2} m^2 \phi^2 + \frac{1}{12} \zeta R \phi^2 \right)~,
\end{equation}
where
$\zeta$ is a parameter from the non-minimal coupling. The corresponding
equation of motion for the scalar field is
\begin{equation}
\label{waveequation1}
\left( \Box - m_{\mbox{\tiny eff}} ^2 \right)\phi=0~,
\end{equation}
where $\Box =
\frac{1}{\sqrt{-g}}\partial_{\alpha}(\sqrt{-g}g^{\alpha\beta}\partial_{\beta})$
is the Laplace-Beltrami operator associated with the metric
(\ref{metric2}) and $m_{\mbox{\tiny eff}} ^2 = m^2 -
\zeta\frac{d(d-2)}{4l^2}$ plays the role of an effective mass for
$\phi$ where $m$ is the mass of the scalar field and $R= -d(d-1) l^{-2}$ is the scalar curvature
\cite{Aros:2002te}. Hence, in this fashion the equation of motion
(\ref{waveequation1}) resembles a field equation for a minimally
coupled scalar field. By means of the following ansatz
\begin{equation}
\phi=\frac{U(r)}{r} Y(\Sigma_{d-2}) e^{-i\omega t}~,
\label{eq:ansatz}
\end{equation}
the radial part of (\ref{waveequation1}), in four dimensions reduces to a
Schrodinger-like equation for a central potential function. Here,
$Y= Y (\Sigma_{d-2})$ is a normalizable harmonic function on
$\sum_{d-2}$ satisfying $\nabla^2Y=-QY$ where $\nabla^2$ is the
Laplace operator and $Q=\left( \frac{d-3}{2}\right)^2 +
\xi^2$, and $\xi$ is any real number \cite{Terras} and
$\omega$ is the frequency of the wave. The radial function $U(r)$
satisfies
\begin{equation}
f(r)\left\{f(r)\frac{d^2}{dr^2}+ f(r)\left[ \frac{d-4}{r} + \frac{f'(r)}{f(r)}\right]\frac{d}{dr}- \frac{d-4}{r^2}f(r)-\frac{f'(r)}{r}-\frac{Q}{r^2}-m_{eff}^2\right\}U(r)+\omega^2\,U(r)=0~,
\end{equation}
where $f'(r) = \frac{d f}{dr}$. By introducing the tortoise coordinate $r_* = r_* (r)$, given by
$dr_*=\frac{dr}{f(r)}$, the latter equation is rewritten as one-dimensional Schrodinger equation,
\begin{equation}
\left[\frac{d^2}{d {r_*}^2} + \omega^2 - V_{\mbox{\tiny eff}}(r)\right]U(r_*)=0~,
\end{equation}
where we can read off immediately the effective potential
\begin{equation}
V_{\mbox{\tiny eff}}(r)=f(r)\left[ m_{\mbox{\tiny eff}}^2 + \frac{Q}{r^2} + \frac{f'(r)}{r}\right]~.
\end{equation}
This potential is depicted in Fig.~(\ref{pot}).
In connection with $f(r)$, explicitly the tortoise coordinate is given by
$r_*=-l \, {\mbox{arctanh}} \left(\frac{r}{l}\right)=-r_+ \,{\mbox{arctanh}} \left(\frac{r}{r_+}\right)$~.
With order to solve analytically the wave equation, the change of variables, $z=1-l^2/r^2$ and
$t=lt$ result useful \cite{Aros:2002te}. By
using the ansatz $\varphi=R(z)Y(\sum)e^{-i\omega t}$, the radial function obeys the following
differential equation
\begin{equation}
\label{Rz}
\left\lbrace z(1-z) \frac{d^2}{d z^2} + \left[ 1 + \left( \frac{d-5}{2}
\right)z \right] \frac{d}{dz} + \left[ \frac{\omega^2}{4z} - \frac{Q}{4} -
\frac{m^2 _{\mbox{\tiny eff}}l^2}{4(1-z)}\right] \right\rbrace R(z) = 0~.
\end{equation}
Assuming that $R(z) = z^\alpha (1-z)^\beta K(z)$, Eq.~(\ref{Rz}) yields
\begin{equation*}
z(1-z) K''(z) + \left[ c - (1+a + b)z \right] K'(z) - ab K(z)  = 0~,
\end{equation*}
whose solution is given in terms of hypergeometric functions \cite{M. Abramowitz}
\begin{equation}
 K(z) = C_1 F(a,b,c,z) + C_2\,z^{1-c} F(a-c+1,b-c+1,2-c,z)~,
\label{Kz}
\end{equation}
with $C_1$ and $C_2$ being constants. The hypergeometric coefficients, $a,b$
and $c$ are defined as follows
\begin{eqnarray*}
\label{a}\
a&=&-\left( \frac{d-3}{4}\right)  + \alpha + \beta_{\pm} + \frac{i}{2}\xi~,
\\
 b&=&-\left( \frac{d-3}{4}\right)  + \alpha + \beta_\pm - \frac{i}{2}\xi~,
\\
 c &=& 1 + 2\alpha~,
\end{eqnarray*}
with $C$ being a non-integer, and
\begin{equation}
\label{alpha}
\alpha= \pm i\frac{\omega}{2}~,\qquad \quad \beta=\beta_\pm =\left( \frac{d-1}{4}
\right)  \pm \frac{1}{2}\sqrt{\left( \frac{d-1}{2}\right)^2 + m_{\mbox{\tiny eff}}^2l^2}~.
\end{equation}
Without loss of generality we choose the negative sign for $\alpha$.
One interesting feature to notice is that the function (\ref{Kz}) has three regular singular points at $z=0$, $z=1$ and $z=\infty$. Then, the solution $R(z)$ is
\begin{equation}
R(z)=C_{1}z^\alpha(1-z)^\beta F_{1}(a,b,c;z)+C_2z^{-\alpha}(1-z)^\beta F_{1}(a-c+1,b-c+1,2-c;z)~.
\end{equation}
Notice that in the neighborhood of the horizon, $z=0$, by using the property $F(a,b,c,0)=1$ \cite{M. Abramowitz},
the function $R(z)$ acquires the form $
R(z)=C_1 e^{\alpha \ln z}+C_2 e^{-\alpha \ln z}$.
Therefore, the scalar field $\phi$ behaves as
\begin{equation}
\phi\sim C_1 e^{-i\omega(t+ \frac{1}{2}\ln z)}+C_2
e^{-i\omega(t- \frac{1}{2}\ln z)}~.
\end{equation}
This expression is quite general as it follows from (\ref{Rz}) and notice that $\phi$ represents
both ingoing and outgoing waves. To be able to
interpret to the scalar field as being only ingoing waves at the horizon the
constant $C_2$ must be eliminated.
The general radial solution with boundary conditions at the horizon can then
be written as
\begin{equation}
\label{horizonsolution4d}\
R(z)=C_1 e^{-i\frac{\omega}{2}\ln z}(1-z)^\beta F_{1}(a,b,c;z)~.
\end{equation}
In order the implement suitable boundary conditions at infinity
($z=1$) for the solution (\ref{horizonsolution4d}) we find
convenient to use the Kummer's relation for the hypergeometric
functions (see for example, \cite{M. Abramowitz})
The radial function is therefore given by
\begin{equation}
\label{Rinfinity}\
R(r) = C_1 \left[ \left(\frac{r_+}{r}\right)^{2\beta}\frac{\Gamma(c)\Gamma(c-a-b)}{\Gamma(c-a)\Gamma(c-b)} + \left(\frac{r_+}{r}\right)^{d -1- 2\beta}\frac{\Gamma(c)\Gamma(a+b-c)}{\Gamma(a)\Gamma(b)}\right]~,
\end{equation}
where we have used the fact that $1-z=\frac{l^2}{r^2}=
\frac{r_+^2}{r^2}$ besides the limit of $R(z)$ when $z\rightarrow1$.


Another way at looking at the radial solution when $r\rightarrow\infty$ at the asymptotic region,
is from the wave equation (\ref{waveequation1})
\begin{equation}\label{radial4daprox}\
R''(r) + \frac{d}{r} R'(r) + \frac{l^2}{r^2} \left(\frac{\omega^2}{r^2} -
\frac{Q}{r^2} - m_{\mbox{\tiny eff}}^2\right)R(r)=0~,
\end{equation}
where we have used the asymptotic behavior of $f(r)$ and $f'(r)$, and the ansatz $\phi=R(r)Y(\sum_{d-2})e^{-i\omega t}$ with $R'(r) = \frac{dR}{dr}$.
The solution to this equation is given in terms of Bessel functions
\cite{M. Abramowitz}
\begin{equation}
R(r)=\left(\frac{\sqrt{A}}{2r}\right)^{\frac{d - 1}{2}}\left[D_1\Gamma{(1-C)}J_{-C}
\left(\frac{\sqrt{A}}{r} \right) + D_2 \Gamma {(1+C)} J_{C} \left( \frac{\sqrt{A}}{r} \right)
\right]~,
\end{equation}
where
\begin{eqnarray*}
A &=& l^2 ( l^2 \omega^2 - Q ) = r_+ ^2 ( r_+ ^2 \omega^2 - Q)~,
\\
 C &=& \frac{1}{2} \sqrt{(d-1)^2 + 4m_{\mbox{\tiny eff}}^2 l^2}~,
\end{eqnarray*}
where $D_1$ and $D_2$ are integration constants. It is
straightforward to simplify the radial solution by using the
expansion of the Bessel function for small arguments, namely
\cite{M. Abramowitz}
\begin{equation*}\label{expansionBessel}\
J_n(x)=\frac{x^n}{2^n\Gamma{(n+1)}}\left\{1-\frac{x^2}{2(2n+2)}+...\right\}~, \qquad
\mbox{for} \quad x \ll 1~.
\end{equation*}
A short calculation shows that the asymptotic radial solution exhibit the polynomial
form
\begin{equation}
R_{\mbox{\tiny asymp}}(r)= \widehat{D}_1 \left( \frac{1}{r} \right)^{\frac{d-1}{2}-C}
+ \widehat{D}_2 \left( \frac{1}{r} \right)^{\frac{d-1}{2}+C}~,
\label{Rasymp1}
\end{equation}
where we have introduced the constants $\widehat{D}_1\equiv D_1\left(\frac{\sqrt{A}}{2}\right)^{\frac{d-1}{2}-C}$ and $\widehat{D}_2\equiv D_2\left(\frac{\sqrt{A}}{2}\right)^{\frac{d-1}{2}+C}$,
also we used
$\frac{\sqrt{A}}{r}<<1$. In Ref.~\cite{Hertog:2004rz}, it was discussed that a scalar field with asymptotic
behavior, similar that  Eq.~(22), generically leads to an unstable state
(the so called, big crunch singularity) which is a clearly sign of
nonlinear instability. However, in order to induce such
instability, this modify the boundary conditions that
scalar field must satisfy at infinity, Ref.~\cite{mezincescu}. Although the
modified boundary conditions preserve the full set of asymptotic
AdS symmetries, and allow for a finite conserved energy to be defined, this energy can be
negative.
We notice, that the imposition of regularity condition on the
radial function (\ref{Rasymp1}) at the infinity implies
$\frac{d-1}{2}-C\geq0$ or $-\frac{(d-1)^2}{4}\leq m_{\mbox{\tiny
eff}}^2l^2\leq 0$. This is in agreement with the condition for
any effective mass in order to have a stable asymptotic AdS spacetime in $d$ dimensions, $m_{\mbox{\tiny eff}}^2l^2\geq-\frac{(d-1)^2}{4}$
\cite{mezincescu} which sets requirements on the nonminimal coupling constant, once
the bare mass of the scalar field and the dimensions are fixed. Besides, $a+b-c=-C$, for $\beta=\beta_-$, and $c-a-b=-C$, for $\beta=\beta_+$. For this reason $C$ can not be an integer, because the gamma function is singular at that point and the regularity conditions are not satisfied.
We now take advantage of the inherent symmetry that the radial solution possesses in the
asymptotic region.  More specifically, we have the freedom to choose the form of the constant $\beta$ since by changing  $\beta_+$ to $\beta_-$  this solution is unchanged. Comparison of Eqs.~(\ref{Rinfinity}) and (\ref{Rasymp1}), regarding $\beta = \beta_-$,  allows us to immediately to
read off the coefficients $\widehat{D}_1$ and $\widehat{D}_2$,
\begin{equation}\label{Dbarra12}\
\widehat{D}_1=C_1 r_{+}^{2\beta_-}\frac{\Gamma(c)\Gamma(c-a-b)}{\Gamma(c-a)\Gamma(c-b)}~,
\quad \qquad \widehat{D}_2=C_1 r_{+}^{d-1 - 2\beta_-}\frac{\Gamma(c)\Gamma(a+b-c)}{\Gamma(a)\Gamma(b)}~.
\end{equation}

\section{Reflection and absorption coefficients. Absorption cross section in a $d$-
dimensional topological massless black hole}

The reflection and absorption coefficients, $\Re$ and $\mathfrak{U}$, respectively,
are defined by
\begin{equation}
\label{reflection4d}\
\Re := \left| \frac{F_{\mbox{\tiny asymp}}^{\mbox{\tiny out}}}{F_{\mbox{\tiny asymp}}^{\mbox{\tiny in}}}
\right| \qquad \qquad \mathfrak{U} := \left|\frac{F_{\mbox{\tiny hor}}^{\mbox{\tiny in}}}{F_{\mbox{\tiny asymp}}^{\mbox{\tiny in}}}\right|~,
\end{equation}
where $F(r)$ is the conserved flux defined by \cite{Satoh:1998ss}
\begin{equation}
\label{flux4d}\
\textit{F}=\frac{\sqrt{-g}g^{rr}}{2i}\left(R^{\ast}
\partial_{r}R - R\partial_{r}R^{\ast}\right)~,
\end{equation}
$R$ being the radial solution of the wave equation (\ref{waveequation1}) and $i$ is the complex
unity and $\ast$ stands for complex conjugation. According to our development the behavior of the
flux $F(r)$ at the horizon is obtained by the introduction of Eq.~(\ref{horizonsolution4d}) into
Eq.~(\ref{flux4d}). Thus, up to an irrelevant factor coming from angular part of the solution,
the flux at the horizon is given by
\begin{equation}
\label{flux4dhorizon}\
\textit{F} _{\mbox{\tiny hor}}^{\mbox{\tiny in}} = - \left| C_{1} \right|^{2} \omega l^{d - 3}~.
\end{equation}
Now, by inserting Eq.~(\ref{Rasymp1}) into Eq.~(\ref{flux4d}), a similar computation leads us to obtain
the flux at the asymptotic region
\begin{equation}
\label{flux4dinfinity}\
\textit{F}_{\mbox{\tiny asymp}} = - i C \left(\frac{1}{l^2 } - \frac{1}{r^2}\right)
\left(\widehat{D}_{2}^*\widehat{D}_{1} - \widehat{D}_{1}^*\widehat{D}_{2} \right)~.
\end{equation}
Nevertheless, the distinction between the ingoing and outgoing fluxes at the asymptotic region is a non
trivial task because the spacetime is asymptotically AdS. In order to characterize the
fluxes we find convenient to split up the coefficients $\widehat{D}_1$ and
$\widehat{D}_2$ in terms of the incoming and outgoing coefficients, $D_{\mbox{\tiny in}}$
and $D_{\mbox{\tiny out}}$, respectively. Making the partition $\widehat{D}_1 = D_{\mbox{\tiny in}}
+ D_{\mbox{\tiny out}}$ and $\widehat{D}_2 = i h (D_{\mbox{\tiny out}} - D_{\mbox{\tiny in}})$
with $h$ being a dimensionless constant which will be assumed to be independent of the energy $\omega$, \cite{Birmingham:1997rj,Kim:1999un, Oh:2008tc, Kao:2009fh}. If we claim physical meaning for the coefficients under study we need specific values for the parameter $h$. We will come back at this point in the last section. In this way  the asymptotic flux Eq.~(\ref{flux4dinfinity}) becomes
\begin{equation}
\textit{F} _{\mbox{\tiny asymp}} \approx \frac{2hC}{l^2}\left( \left|D_{\mbox{\tiny in}}
\right|^{2} - \left|D_{\mbox{\tiny out}} \right|^{2}\right)~.
\end{equation}
Therefore, the coefficients Eq.~(\ref{reflection4d}) are given by
\begin{eqnarray}
\label{r4d}
\Re &=& \frac{\left|D_{\mbox{\tiny out}} \right|^2}{\left|D_{\mbox{\tiny in}}\right|^2}~,
\\
\mathfrak{U}&=&\frac{\omega l^{d-1} \left|C_{1}\right|^2}{2 \left| h \right| C \left|D_{\mbox{\tiny
in}}\right|^2}~,
\end{eqnarray}
where the coefficients $D_{\mbox{\tiny in}}$ and $D_{\mbox{\tiny out}}$, are expressed as
\begin{eqnarray}
\label{D14d}\ D_{\mbox{\tiny in}}&=& \frac{C_1}{2}
\left[r_{+}^{2\beta_-}\frac{\Gamma(c)\Gamma(c-a-b)}{\Gamma(c-a)\Gamma(c-b)}
+
\frac{i}{h}r_{+}^{d-1-2\beta_-}\frac{\Gamma(c)\Gamma(a+b-c)}{\Gamma(a)\Gamma(b)}\right]~,
\\
\label{D24d}\ D_{\mbox{\tiny out}}&=& \frac{C_1}{2}
\left[r_{+}^{2\beta_-}\frac{\Gamma(c)
\Gamma(c-a-b)}{\Gamma(c-a)\Gamma(c-b)} -
\frac{i}{h}r_{+}^{d-1-2\beta_-}\frac{\Gamma(c)\Gamma(a+b-c)}{\Gamma(a)\Gamma(b)}\right]~.
\end{eqnarray}

On the other hand, the absorption cross section, or greybody factor $\sigma_{\mbox{\tiny abs}}$, is given by
\begin{equation}
\sigma_{\mbox{\tiny abs}} = \frac{\mathfrak{U}}{\omega}=\frac{l^{d-1}\left|C_{1}\right|^2 }{2\left|h\right|
C \left|D_{\mbox{\tiny in}}\right|^2}~.
\end{equation}

\begin{figure}
\includegraphics[width=4.0in,angle=0,clip=true]{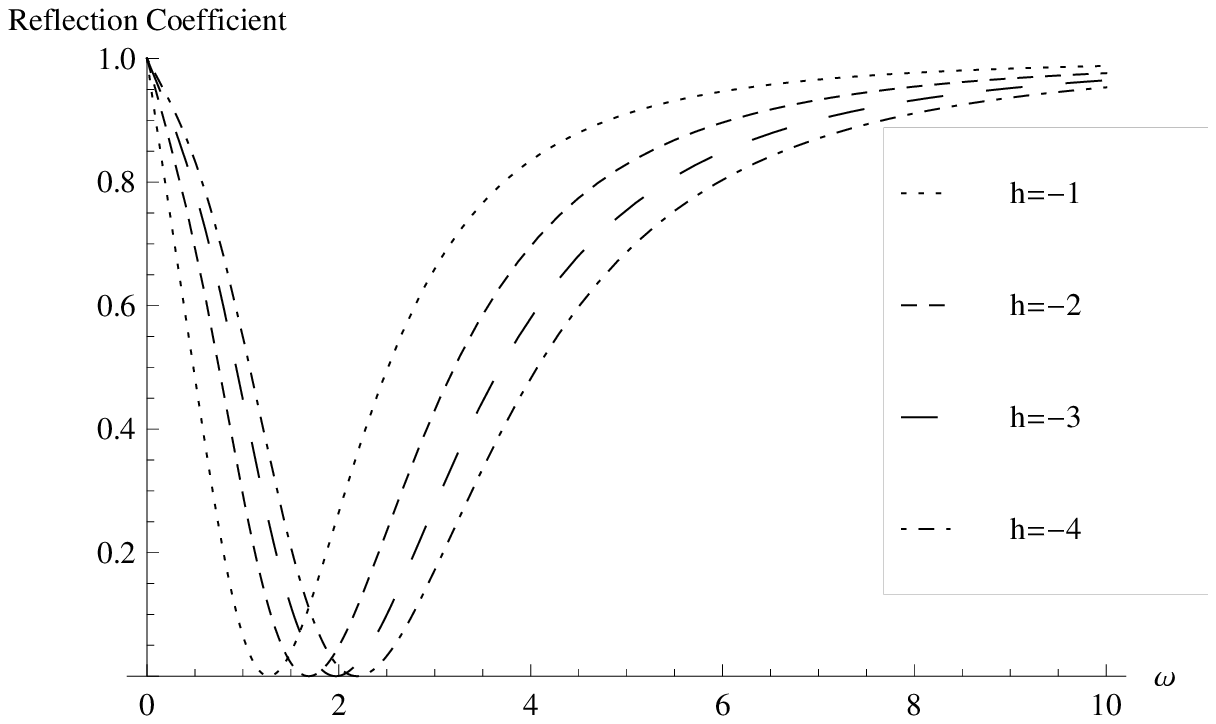}
\caption{Reflection coefficient v/s $\omega$; $d=4$, $m_{\mbox{\tiny eff}}^2l^2=0$, $l=1$ and $\xi=0$.}
\label{ReflectionCoefficientTMBH4dh}
\end{figure}
\begin{figure}
\includegraphics[width=4.0in,angle=0,clip=true]{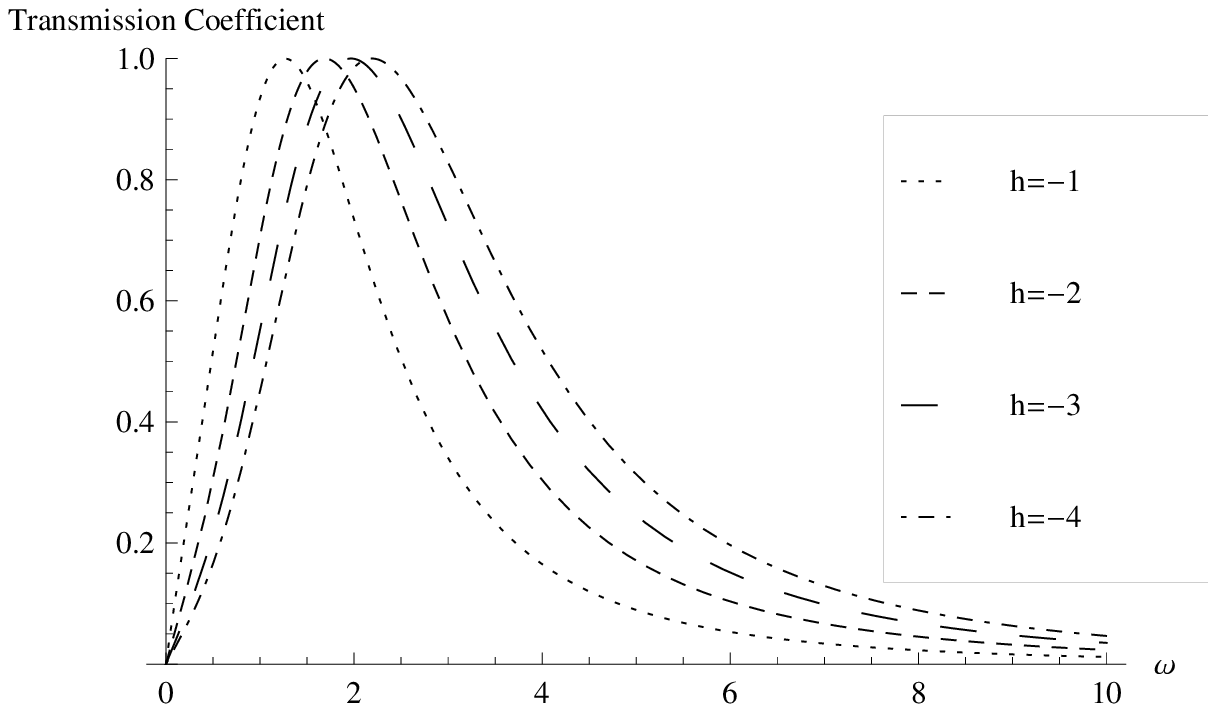}
\caption{Transmission coefficient v/s $\omega$; $d=4$, $m_{\mbox{\tiny eff}}^2l^2=0$, $l=1$ and $\xi=0$.}
\label{TransmissionCoefficientTMBH4dh}
\end{figure}
\begin{figure}
\includegraphics[width=4.0in,angle=0,clip=true]{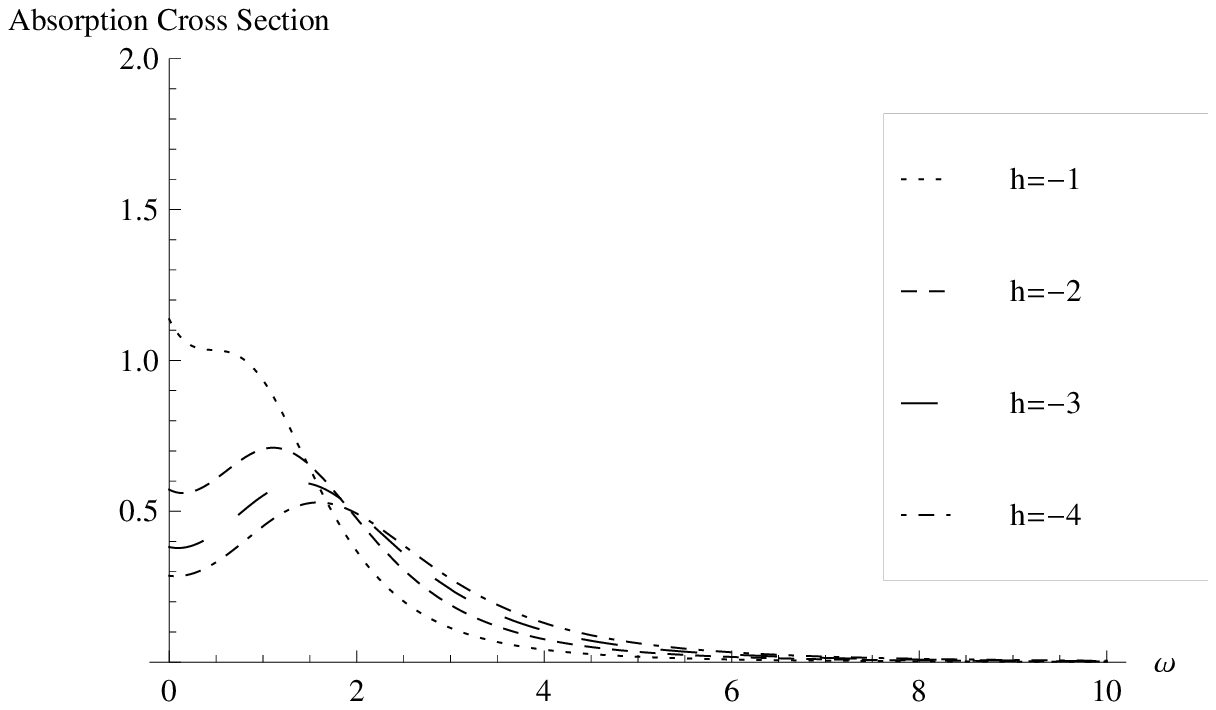}
\caption{Absorption Cross Section v/s $\omega$; $d=4$, $m_{\mbox{\tiny eff}}^2l^2=0$, $l=1$, and $\xi=0$.}
\label{AbsorptionCrossSectionTMBH4dh}
\end{figure}
\begin{figure}
\includegraphics[width=4.0in,angle=0,clip=true]{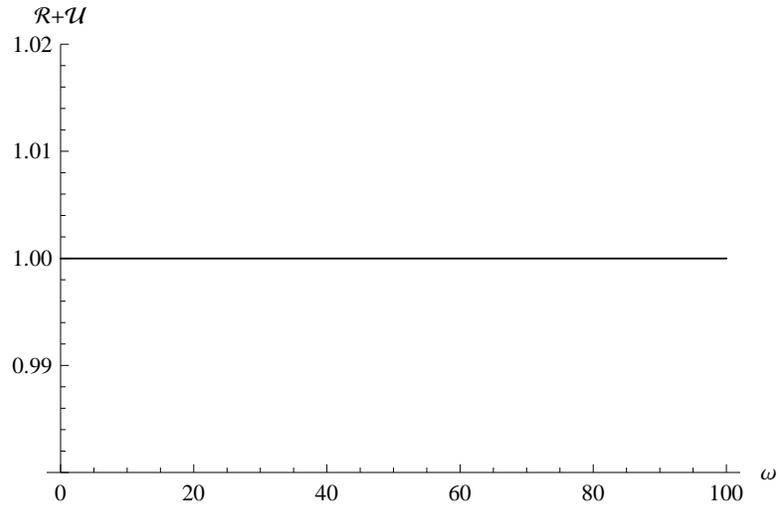}
\caption{Physical conditions $\Re+\mathfrak{U}$ is plotted against $\omega$; $d=4$, $m_{\mbox{\tiny eff}}^2l^2=0$, $l=1$, $\xi=0$ and $h=-1,-2,-3,-4$. This figure shows us the physical requirement is satisfied for negative values of parameter $h$.}
\label{ConditionTMBH4dh}
\end{figure}
\begin{figure}
\includegraphics[width=4.0in,angle=0,clip=true]{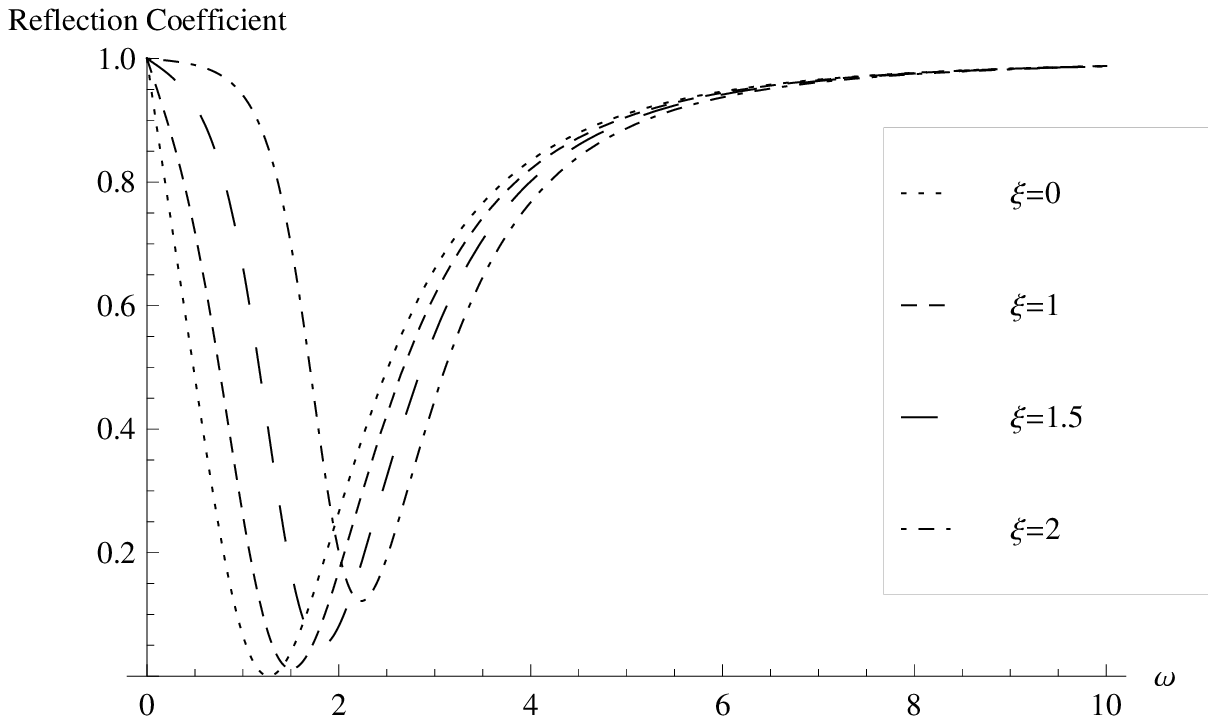}
\caption{Reflection coefficient v/s $\omega$; $d=4$, $m_{\mbox{\tiny eff}}^2l^2=0$, $l=1$ and $h=-1$.}
\label{ReflectionCoefficientTMBH4d}
\end{figure}
\begin{figure}
\includegraphics[width=4.0in,angle=0,clip=true]{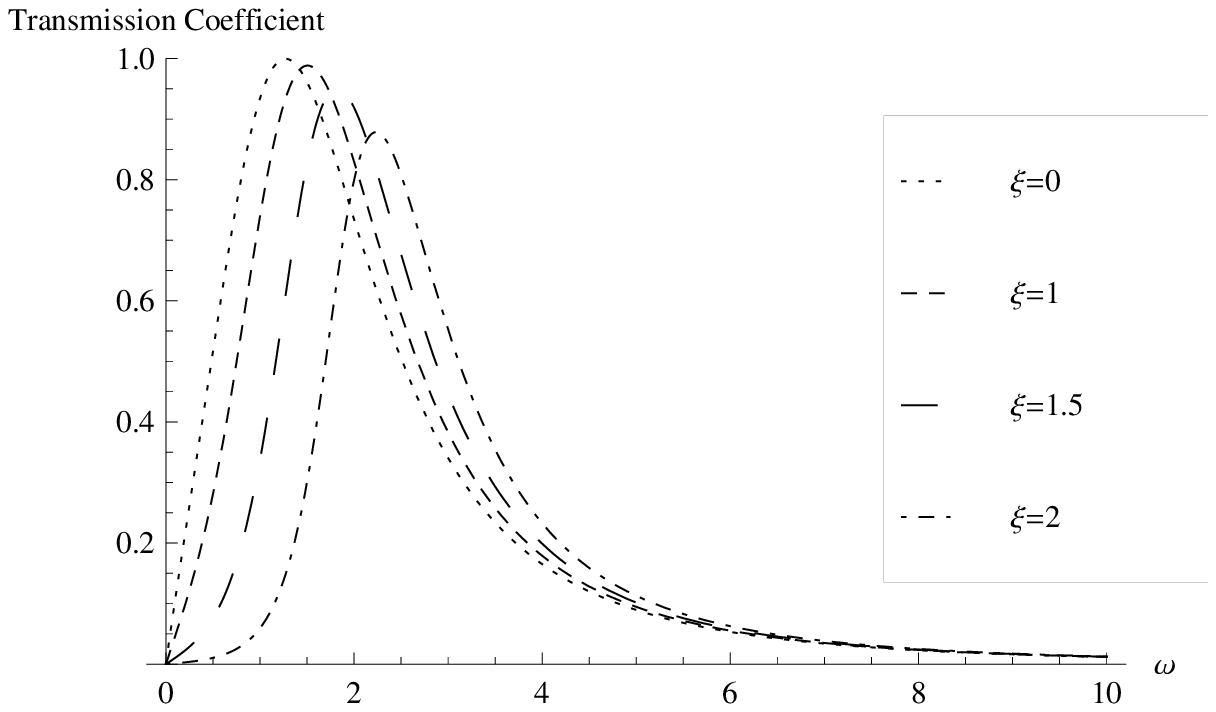}
\caption{Transmission coefficient v/s $\omega$; $d=4$, $m_{\mbox{\tiny eff}}^2l^2=0$, $l=1$ and $h=-1$.}
\label{TransmissionCoefficientTMBH4d}
\end{figure}
\begin{figure}
\includegraphics[width=4.0in,angle=0,clip=true]{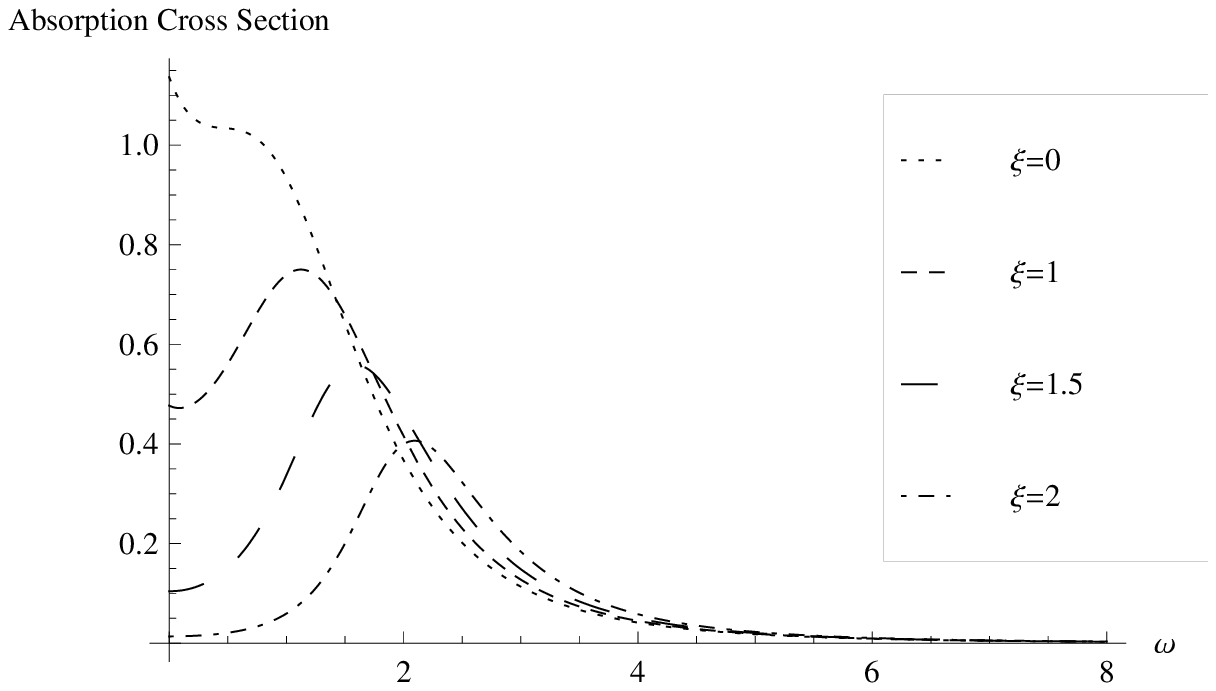}
\caption{Absorption Cross Section v/s $\omega$; $d=4$, $m_{\mbox{\tiny eff}}^2l^2=0$, $l=1$ and $h=-1$.}
\label{AbsorptionCrossSectionTMBH4d}
\end{figure}
\begin{figure}
\includegraphics[width=4.0in,angle=0,clip=true]{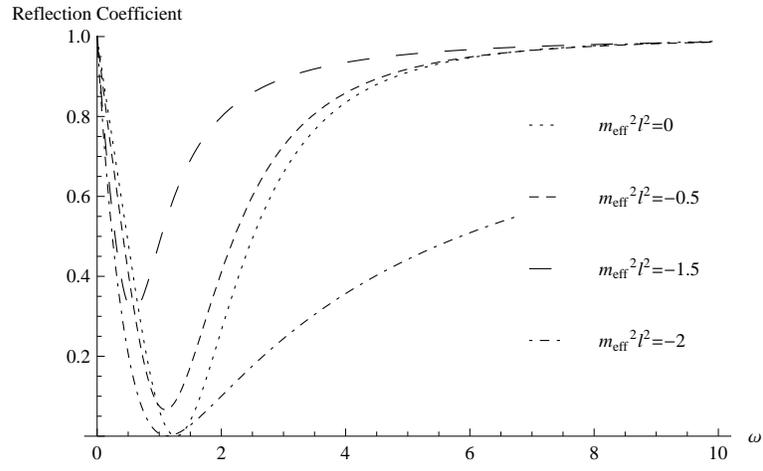}
\caption{Reflection coefficient v/s $\omega$; $m_{\mbox{\tiny eff}}^2l^2=-2,-1.5,-0.5,0$, $l=1$, $h=-1$, $\xi=0$ and $d=4$.}
\label{ReflectionCoefficientTMBH4dm}
\end{figure}
\begin{figure}
\includegraphics[width=4.0in,angle=0,clip=true]{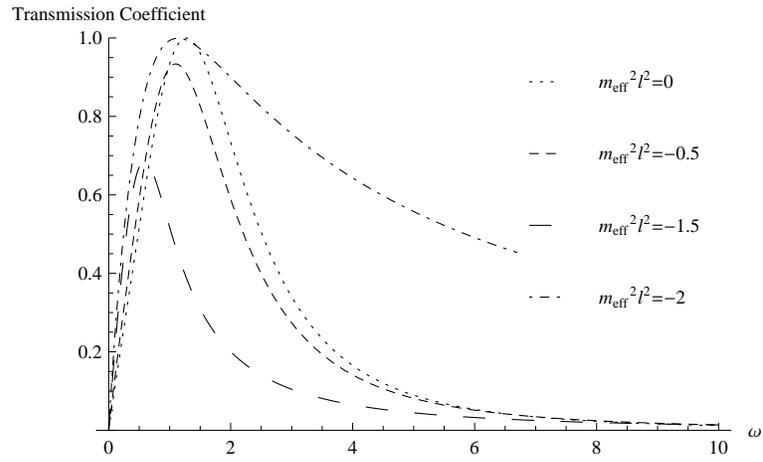}
\caption{Transmission coefficient v/s $\omega$; $m_{\mbox{\tiny eff}}^2l^2=-2,-1.5,-0.5,0$, $l=1$, $h=-1$, $\xi=0$ and $d=4$.}
\label{TransmissionCoefficientTMBH4dm}
\end{figure}
\begin{figure}
\includegraphics[width=4.0in,angle=0,clip=true]{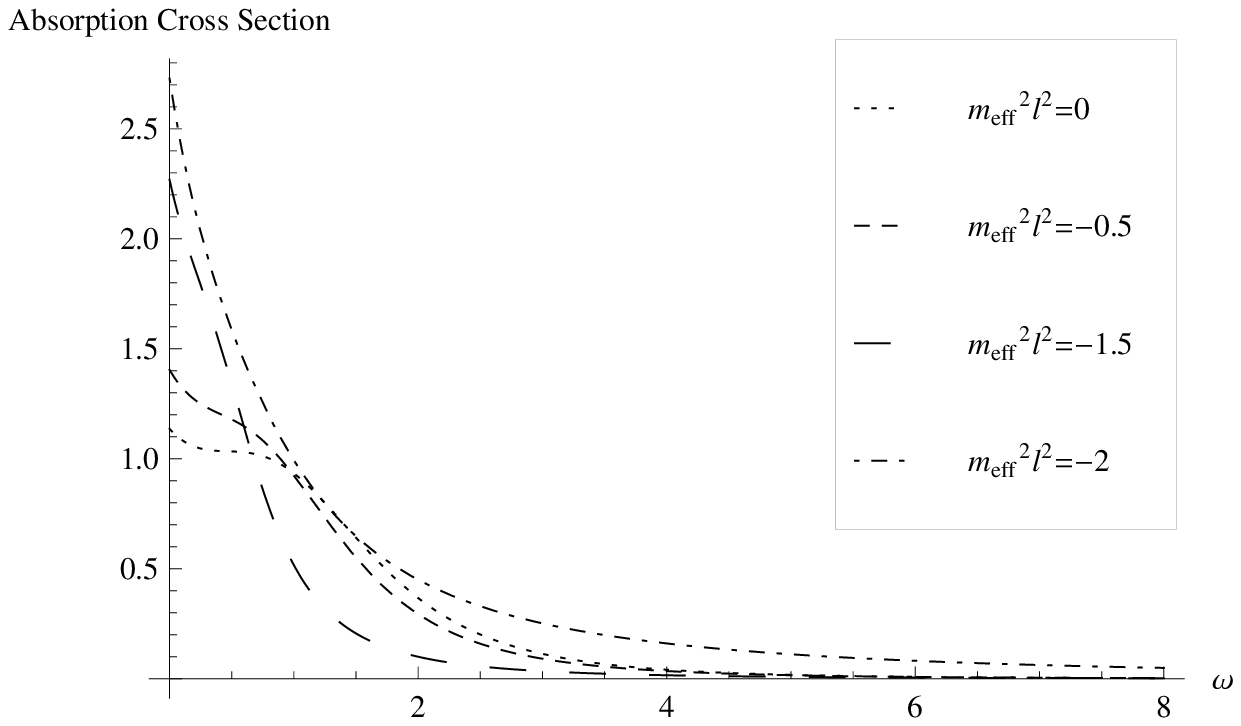}
\caption{Absorption Cross Section v/s $\omega$; $m_{\mbox{\tiny eff}}^2l^2=-2,-1.5,-0.5,0$, $l=1$, $h=-1$, $\xi=0$ and $d=4$.}
\label{AbsorptionCrossSectionTMBH4dm}
\end{figure}

\subsection{4-dimensional case}

If we restrict our general results developed above to the
$4$-dimensional case, we minimize the amount of formalism. We
focus mainly on the radial solutions for the wave equation
(\ref{waveequation1}) in order to obtain the scalar perturbations
on a $4$-dimensional massless black hole. For such a case ($d=4$),
from (\ref{Rinfinity}) and (\ref{Rasymp1}) we have immediately the
radial solutions with their corresponding behaviors
\begin{equation}\label{Rinfinity4d}\
R(r)=C_1 \left[ \left(\frac{r_+}{r}\right)^{2\beta}\frac{\Gamma(c)\Gamma(c-a-b)}{\Gamma(c-a)\Gamma(c-b)} + \left(\frac{r_+}{r}\right)^{3- 2\beta}\frac{\Gamma(c)\Gamma(a+b-c)}{\Gamma(a)\Gamma(b)}\right]~,
\end{equation}
\begin{equation}\label{Rasymp4d}
R_{\mbox{\tiny asymp}}(r)= \widehat{D}_1 \left( \frac{1}{r} \right)^{\frac{3}{2}-C}
+ \widehat{D}_2 \left( \frac{1}{r} \right)^{\frac{3}{2}+C}~.
\end{equation}

Note that the radial function (\ref{Rasymp4d}) satisfies the regularity condition at
infinite if $\frac{3}{2}-C\geq0$ or $-\frac{9}{4}\leq
m_{\mbox{\tiny eff}}^2l^2\leq 0$. This condition is in agreement
with the Breitenlohner-Freedman bound for the positivity of energy
in global AdS$_4$, \cite{Breitenlohner:1982jf,
Breitenlohner:1982bm}, $m_{\mbox{\tiny
eff}}^2l^2\geq-\frac{9}{4}$, which sets requirements on the nonminimal coupling constant once
the bare mass of the scalar field and the dimensions are fixed. Besides, $a+b-c=-C$, for $\beta=\beta_-$, and $c-a-b=-C$, for $\beta=\beta_+$.  This is the reason why $C$ can not be an integer because the gamma function is singular at that point and the regularity conditions are not fulfilled. On the other hand, the solution given by Eq.~(35) is symmetric by changing $\beta_-$ to $\beta_+$. This leads us to consider $\beta_-$ without loss of generality. The comparison between radial solutions becomes
\begin{equation}\label{Dbarra}\
\widehat{D}_1=C_1 r_{+}^{2\beta_-}\frac{\Gamma(c)\Gamma(c-a-b)}{\Gamma(c-a)\Gamma(c-b)}~,
\quad \qquad \widehat{D}_2=C_1 r_{+}^{3- 2\beta_-}\frac{\Gamma(c)\Gamma(a+b-c)}{\Gamma(a)\Gamma(b)}~,
\end{equation}
with $C_1$ being an integration constant. Now, with respect to the reflection and absorption coefficients in four dimensions we have straightforwardly
\begin{equation}
 \label{R-and-U}
\Re = \frac{\left|D_{\mbox{\tiny out}} \right|^2}{\left|D_{\mbox{\tiny in}}\right|^2}
\qquad {\mbox{and}} \qquad
\mathfrak{U}=\frac{\omega l^{3} \left|C_{1}\right|^2}{2 \left| h \right| C \left|D_{\mbox{\tiny
in}}\right|^2}~,
\end{equation}
where the coefficients are given by (\ref{D14d}) and (\ref{D24d}) specialized to $d=4$.
In closing this section, we have the corresponding greybody factor
\begin{equation}
 \sigma_{\mbox{\tiny abs}} = \frac{\mathfrak{U}}{\omega}=\frac{l^{3}\left|C_{1}\right|^2 }{2\left|h\right|
C \left|D_{\mbox{\tiny in}}\right|^2}~.
\end{equation}

A numerical analysis of the coefficients is depicted in Figs.~(2-11).

\section{Discussions and comments}

In this paper we have computed the greybody factors, the reflection and transmission coefficients for topological massless black holes in arbitrary dimensions.
The involved physical content by the coefficients is more suitable
viewed when we evaluate them numerically in four dimensions.
Our numerical analysis needed of specific choices for the aforementioned constant $h$.
To this respect we made allowance for the currently accepted
discussion for the selection of the parameter $h$.
On the one hand, the constant $h$ can be chosen conveniently in
such a manner that the absorption cross section can be expressed
by the area of horizon in the zero-frequency limit
\cite{Birmingham:1997rj, Das:1996we}.
On the other hand, it can be chosen also, in order to obtain the
correct value of the Hawking temperature \cite {Kim:1999un}, besides
that of completeness where is necessary to assure that the sum of
the reflection and transmission coefficients becomes the unity
\cite {Kim:2004sf}.
In addition, it has been reported that this freedom in the choice of
$h$ as a numerical factor is usually set up by imposing appropriate
physical conditions \cite{Oh:2008tc}.
Thus, according to our concern we proceeded to choose the parameter $h$ in such way that either the values of the
greybody factors, reflection and transmission coefficients represent
an acceptable physical situation.
In this sense we employed $h$ as a free parameter and we use it to plot
the reflection coefficient (see Fig.~(\ref{ReflectionCoefficientTMBH4dh})),
transmission coefficient (see Fig.~(\ref{TransmissionCoefficientTMBH4dh}))
and the greybody factors (see Fig.~(\ref{AbsorptionCrossSectionTMBH4dh})),
for some values of $m_{\mbox{\tiny eff}}, l$ and $\xi$.
We found that negative values for the parameter $h$ provide physical meaning
whereas in the positive case some of the coefficients under study become divergent. We observed also that the parameter $h$ must be bounded.
It results smaller than zero and greater than some other value, such that
the absorption cross section or the greybody factor become real in the
zero-frequency limit. Likewise, in this range
the greybody factor is such that this coefficient is increasing if
the parameter $h$ is increasing, (see Fig.~(\ref{AbsorptionCrossSectionTMBH4dh})).
Besides, for completeness in our description we have plotted the condition $\Re+\mathfrak{U}$, Fig.~(\ref{ConditionTMBH4dh}), for $m_{\mbox{\tiny eff}}^2l^2=0$,
$l=1$, $\xi=0$ and $h=-1,-2,-3,-4$. This condition is satisfied being
equal to the unity. On the other hand, as mentioned previously, in four
dimensions, $-\frac{9}{4}\leq m_{\mbox{\tiny eff}}^2l^2\leq 0$ and $C$ can
not be an integer or equivalently $m_{\mbox{\tiny eff}}^2l^2\neq-\frac{9}{4},
-\frac{5}{4}$, because for these values the regularity conditions are not fulfilled.
Therefore, we consider without loss of generality, $m_{\mbox{\tiny eff}}^2l^2=0$, $l=1$ and $h=-1$, as a fixed parameter useful to analyze the behavior of all the
coefficients in four dimensions.
Along these lines of reasoning, for the values $\xi=0,1,1.5,2$ we have in Figs.~(\ref{ReflectionCoefficientTMBH4d}), (\ref{TransmissionCoefficientTMBH4d}) and (\ref{AbsorptionCrossSectionTMBH4d}) the reflection and transmission coefficients as
well as the greybody factors, respectively. We point out the existence of a minimum and maximum point for the reflection and
transmission coefficients. We note further that for these coefficients we have two
branches. In the reflection case, one of them is decreasing for low frequencies and
the other one is increasing. For the transmission case the behavior
is contrary to the latter, increasing and then decreasing. This offer to us with
some valuable insight about the existence of one optimal frequency to transfer
energy out of the bulk. Additionally, we found that there is a range of values of $\xi$ that contribute to the greybody factor (see Fig.~(\ref{AbsorptionCrossSectionTMBH4d})) in the zero-frequency limit, contrary to the case studied by Dass, Gibbons and Matur \cite{Das:1996we}, where the mode with lowest angular momentum contribute at the absorption cross section, in that limit. This reason does not allow to fix the value of $h$. We would like to mention further that we notice the effect of the $m_{\mbox{\tiny eff}}$ on the coefficients, (see Figs~(\ref{ReflectionCoefficientTMBH4dm}), (\ref{TransmissionCoefficientTMBH4dm}) and (\ref{AbsorptionCrossSectionTMBH4dm})), where the absorption cross section decreases in the zero-frequency limit when $m_{\mbox{\tiny eff}}$ increases. Finally, we mention that the case $\mu\neq 0$ is a rather involved computation and we are currently working on this point. This will be reported elsewhere.

\section*{Acknowledgments}

This work was supported by COMISION NACIONAL DE CIENCIAS Y
TECNOLOGIA through FONDECYT \ Grant 7080205 (CC, ER, JS), 1090613 (JS). This
work was also partially supported by PUCV DII (JS). ER acknowledges partial support from grants PROMEP (CA: Investigaci\'{o}n y Ense\~{n}anza de la F\'{i}sica, and Promep Redes Tem\'{a}ticas). CC acknowledges partial support from PROMEP Grant 103.5/08/3228.
PG was supported by Direcci\'{o}n de Estudios Avanzados PUCV. The authors
acknowledge the referee for useful suggestions in order to improve
the presentation of the results of this paper.


\end{document}